\documentclass[12pt,fleqn]{article} \textheight=9in
\newcommand{\ra}{\rightarrow}
\newcommand{\bra}{\langle} \newcommand{\ket}{\rangle}
\newcommand{\be}{\begin{equation}}
\newcommand{\ee}{\end{equation}}
\newcommand{\bea}{\begin{eqnarray}}
\newcommand{\eea}{\end{eqnarray}}
\newcommand{\eps}{\epsilon}

\newcommand{\ep}{\quad {\vrule height 10pt width 10pt depth 2pt}}

\newcommand{\grintl}{[\kern-.18em [}
\newcommand{\grintr}{]\kern-.18em ]}
\newcommand{\ds}{\displaystyle}
\newtheorem{lem}{Lemma}[section]
\newtheorem{prop}{Proposition}[section]
\newtheorem{thm}{Theorem}[section]

\def\smallR{\hbox{\scriptsize I\kern-.23em{R}}}
\def\R{\hbox{$\mit I$\kern-.33em$\mit R$}}
\def\C{\hbox{$\mit I$\kern-.6em$\mit C$}}
\def\un{\hbox{$\mit I$\kern-.77em$\mit I$}}
\def\0{\hbox{$\mit I$\kern-.70em$\mit O$}}
\def\r{I\kern-.277em R}

\def\N{\mbox{\bf N}}

\begin{document}

\title{Time Development of Exponentially
Small Non-Adiabatic Transitions
}
\author{George A. Hagedorn\thanks{Partially
Supported by National Science Foundation
Grants DMS--0071692 and DMS--0303586.}\\
Department of Mathematics and\\
Center for Statistical Mechanics and Mathematical Physics\\
Virginia Polytechnic Institute and State University\\
Blacksburg, Virginia 24061-0123, U.S.A.\\[15pt]
\and
Alain Joye\\
Institut Fourier\\ Unit\'e Mixte de Recherche CNRS-UJF 5582\\
Universit\'e de Grenoble I\\
BP 74\\
F--38402 Saint Martin d'H\`eres Cedex, France}

\date{}
\maketitle

\vskip 1.5cm
\begin{abstract}
Optimal truncations of asymptotic expansions are known to yield 
approximations to adiabatic quantum evolutions that are accurate up 
to exponentially small errors.
In this paper, we rigorously determine the leading order 
non--adiabatic corrections to these approximations for a
particular family of two--level analytic Hamiltonian functions. 
Our results capture the time development
of the exponentially small transition  
that takes place between optimal states
by means of a particular switching function.
Our results confirm the physics predictions of
Sir Michael Berry in the sense that
the switching function for this family of Hamiltonians
has the form that he argues is universal.
\end{abstract}

\newpage

\section{Introduction}\label{intro}
\setcounter{equation}{0}

\vskip .5cm
The adiabatic approximation in quantum mechanics asymptotically
describes solutions to the time dependent Schr\"odinger 
equation when the Hamiltonian of the system is a slowly
varying function time.
After a rescaling of the time variable, the adiabatic approximation
describes the small $\eps$ behavior of solutions to 
the Schr\"odinger equation
\be\label{SE} i\,\eps\,\frac{\partial\psi}{\partial t}\ =\ H(t)\,\psi.
\ee

In the simplest non-trivial situation,
$\{H(t)\}_{t\in\smallR}$ is a family of $2 \times 2$ Hermitian 
matrices that depends smoothly on $t$, and whose eigenvalues
$E_1(t)$ and  $E_2(t)$ are separated by a minimal gap
$E_2(t)-E_1(t)\,>\,g\,>\,0$ for all $t\,\in\,\R$.

To discuss scattering transition amplitudes, we
also assume that $H(t)$ approaches limits as $t$ tends to plus or
minus infinity. We let
$\Phi_j(t)$, $j=1,2$ be smooth normalized instantaneous eigenstates
associated with $E_j(t)$, respectively.
Then the transition amplitude ${\cal A}(\eps)$ across 
the gap between the asymptotic eigenstates is defined as 
\be
{\cal A}(\eps)\ =\ \lim_{t_0\ra-\infty \atop t_1\ra +\infty }\ 
|\,\bra\,\Phi_2(t_1),\ U_{\eps}(t_1,\,t_0)\,\Phi_1(t_0)\,\ket\,|,
\ee
where $U_\eps(t_1,\,t_0)$ denotes the evolution operator corresponding to 
(\ref{SE}). The adiabatic theorem of quantum mechanics, \cite{bf},
asserts that 
${\cal A}(\eps)\,=\,O(\eps)$, so the transition probability
${\cal A}(\eps)^2$ is of order $O(\eps^2)$.

\vskip .5cm
If the Hamiltonian is an analytic function of time, 
the transition amplitude is much smaller.
Long ago, Zener \cite{z} considered
a specific real symmetric two-level system, that
had an exponentially small transition ${\cal A}(\eps)$ 
as $\eps\ra 0$. 
Generalizations of Zener's result to analytic real symmetric two-level 
Hamiltonians with a gaps in their spectra were then proposed
in the physics literature. Formulas for
${\cal A}(\eps)$ of the form
\be\label{trapro}
{\cal A}(\eps)\ \simeq\ e^{-\gamma/\eps},
\ \ \ \mbox{ as } \ \ \ \eps\ra 0,
\ee
with $\gamma >0$ that applied more generally were obtained, {\it e.g.},
in \cite{lan}, \cite{d}, \cite{hp}.
The decay rate $\gamma$ was essentially determined
by complex crossing points, {\it i.e.}, the points in the complex 
$t$--plane where the analytic continuations of the eigenvalues coincided.
Later on, papers \cite{b2} and \cite{jkp} recognized independently
that non-trivial prefactors $G$, with $|G|\neq 1$ could be present
for general 
Hermitian two-level Hamiltonians to yield the general formula
\be\label{trapro1}
{\cal A}(\eps)\ \simeq\ |G|\ e^{-\gamma/\eps},
\ \ \ \mbox{ as } \ \ \ \eps\ra 0.
\ee 
Also, \cite{bl} and \cite{j0} pointed out independently
that certain complex degeneracies could lead to the same formula
in the real symmetric case.
Formulas (\ref{trapro}) and (\ref{trapro1}) are variants of
the well--known Landau-Zener formula that has been widely used
in many areas of atomic and molecular physics.

\vskip .5cm
The goal of this paper is to obtain more precise results for a family
of two--level systems. For all times $t$,
we construct an approximate solution to (\ref{SE}) 
that is accurate up to errors of order
$\exp(-\gamma/\eps)\,\eps^{\mu}$, for some $\mu>0$.
This captures the transition process
which is of order $\exp(-\gamma/\eps)$. 

\vskip .5cm
Explicitly, let $E>0$ and $\delta>0$,
and consider the Hamiltonian function
\be\label{H} H(t)\ =\
\frac E{2\,\sqrt{t^2+\delta^2}}\
\left(\,\begin{array}{cc}\delta&t\\t&-\delta\end{array}\,\right)
\ee
whose eigenvalues are $\pm E/2$ for every $t$. This Hamiltonian 
can be viewed as the familiar Landau-Zener Hamiltonian
modified to keep its eigenvalues constant.
In our case, the notion of eigenvalue 
crossing point is replaced by the singularities of the Hamiltonian 
itself at $t\,=\,\pm\,i\,\delta$. These points govern 
the transitions between the two levels. Let $\Phi_1(t)$ and $\Phi_2(t)$
be smooth, normalized real eigenvectors corresponding to $ -E/2$
and $E/2$ respectively.

\vskip .5cm
\begin{thm}\label{supertheorem}
Let $H(t)$ be given by (\ref{H}) and let $0<\mu<1/2$.  Then:

\vskip .3cm
1) There exist vectors
$\chi_1(\eps,\,t)$ and  $\chi_2(\eps,\,t)$ that satisfy the
Schr\"odinger equation (\ref{SE}) up to errors of order $e^{-E\,\delta/\eps}$
and correspond to the eigenstates $\Phi_1(t)$, and $\Phi_2(t)$ in the sense 
that
\be
\lim_{|t|\ra \infty}\ |\,\bra\,\Phi_j(t),\,\chi_j(\eps,\,t)\,\ket\,|\ =\
1\,+\ O(e^{-E\,\delta/\eps}\,\eps^{\mu}).
\ee
Moreover, the set $\{\chi_j(\eps,\,t)\}_{j=1,2}$ is orthonormal up to 
errors of order $e^{-E\,\delta/\eps}\,\eps^{\mu}$.

\vskip .3cm
2) The Schr\"odinger equation has solutions $\Psi_j(\eps,\,t)$, $j=1,2$ such 
that uniformly in $t\in\R$ as $\eps\ra 0$, 
\bea\label{wow1}\nonumber
\Psi_1(\eps,\,t)\ =\ \chi_1(\eps,\,t)\ -\ {\sqrt 2}\
e^{-E\,\delta/\eps}\ \frac 12\
\left\{\,
\mbox{\rm erf}\left(\,
\sqrt{\frac{E}{2\,\delta\,\eps}}\ t\,\right)
\,+\,1\,\right\}\
\chi_2(\eps,\,t)
+\ O(e^{-E\,\delta/\eps}\,\eps^{\mu}),
\eea
and
\bea\nonumber\label{wow2}
\Psi_2(\eps,\,t)\ =\ \chi_2(\eps,\,t)\ +\ {\sqrt 2} 
\ e^{-E\,\delta/\eps}\ \frac 12\
\left\{\,
\mbox{\rm erf}\left(\,
\sqrt{\frac{E}{2\,\delta\,\eps}}\ t\,\right)
\,+\,1\,\right\}\
\chi_1(\eps,\,t)
\ +\ O(e^{-E\,\delta/\eps}\,\eps^{\mu}).
\eea
\end{thm}

\vskip .5cm
\noindent {\bf Remarks}\newline
{\bf 0.} Recall that the function erf is defined by 
\be\label{deferf}
  \mbox{\rm erf}(x)\ =\ \frac{2}{\sqrt{\pi}}\ \int_0^x\ e^{-y^2}\  
dy\ =\ \frac{2}{\sqrt{\pi}}\
\int_{-\infty}^x\ e^{-y^2}\ dy\ -\ 1\ \in\ [-1,\,1].
\ee
{\bf 1.} The vectors $\chi_j(\eps,\,t)$, $j=1,2$, are constructed 
as approximate solutions to (\ref{SE}) obtained by means of 
optimal truncation of asymptotic expansions of actual solutions. 
As $t\ra -\infty$ they are asymptotic to the instantaneous eigenvectors 
$\Phi_j(t)$ of $H(t)$, up to a phase. We call them optimal adiabatic 
states. See (\ref{opti1}), (\ref{opti2}) and (\ref{changevar}).
\newline
{\bf 2.}  The transition mechanism between optimal 
adiabatic states takes place from the value zero to the value 
${\sqrt 2}\ e^{-E\,\delta/\eps}$  
in a smooth monotonic way described by the switching function
(erf+1)/2, on a time
scale of order $\sqrt{\eps}$.  By contrast, the transition between
instantaneous eigenstates of the Hamiltonian displays oscillations 
of order $\eps$ for
any finite time to eventually reach its exponentially small value 
only at $t=\infty$. In that sense also, the
vectors $\chi_j(\eps,\,t)$ are optimal. 
\newline
{\bf 3.} As the optimal adiabatic states and eigenstates
essentially coincide at $t=\pm \infty$, the transition amplitude equals 
${\cal A(\eps)}\ \simeq\ {\sqrt 2}\,e^{-E\,\delta/\eps}$,
up to errors of order 
$e^{-E\,\delta/\eps}\,\eps^{\mu}$.
\newline
{\bf 4.} The parameters $E$ and $\delta$ play somewhat different roles.
If we fix $E$ and decrease $\delta$, the
singularity of $H(t)$ approaches the real axis. The transition amplitude 
${\sqrt 2}\,e^{-E\,\delta/\eps}$ increases whereas
the time it takes to accomplish the transition decreases as
$\sqrt{2\,\delta\,\eps/E}$.
If instead we fix $\delta$ and let $E$ decrease, the gap between the
eigenvalues decreases.
The transition amplitude increases as well, whereas the typical 
time of the transition now increases.
\newline
{\bf 5.} Our results allow us to control the evolution operator
$U_{\eps}(t,\,s)$
associated with (\ref{SE}) up to errors of order 
$e^{-E\,\delta/\eps}\,\eps^{\mu}$, for any time interval $[s,\,t]$.
\newline
{\bf 6.} Further comments concerning the relevance of the Hamiltonian
(\ref{H}) are presented at the
end of this section.

\vskip .5cm
We now put our results in perspective by describing previous work on 
exponential asymptotics for the adiabatic approximation.

Rigorous computations of the behavior of the exponentially
small quantity ${\cal A}(\eps)$
for two-level systems, or generalizations of this typical
setting, were provided relatively late in \cite{jkp}, \cite{jp2}, \cite{j0}, 
\cite{j}, \cite{jp3}, \cite{j1}, \cite{jp4}. 
Although we shall not use the technique in this paper, 
let us briefly describe
the mechanism that is typically
used to get the asymptotics leading to the exponentially
small quantity ${\cal A}(\eps)$ in these papers.
It involves deforming integration paths from
the real axis to the complex plane $t$--plane until they 
reach a (non--real) crossing point. 
Crossing points provide singularities where significant 
transitions take place, but their lying away in the complex plane makes these 
transitions exponentially small due to the presence of dynamical phases 
$\exp(-i\int_0^tE_j(s)\,ds/\eps)$ whose exponents
acquire a non--zero real part along the path.
With this approach, the link 
with the
initial problem posed on the real axis is possible only at infinity.
One does not learn about the dynamics of the transition. 
We note 
that in a more general framework where ${\cal A}(\eps)$
represents the transition between two isolated bands of the spectrum
for general (unbounded) analytic Hamiltonians, exponential bounds on 
${\cal A}(\eps)$ were obtained by suitable adaptations of this method in 
\cite{jp1}, \cite{js}. See also \cite{m}, \cite{sj} for similar results
using the pseudo-differential operator machinery.

\vskip .5cm
The other successful method used to construct precise 
approximations of solutions to (\ref{SE}) uses optimal truncation
of asymptotic expansions. With sufficiently sharp estimates of the
errors, one can prove exponential accuracy. 
Under appropriate analyticity assumptions, one typically proves
that the error committed by retaining $n$ terms in the
asymptotic expansion is of order $n!\,\eps^n$. This error is minimized
by choosing $n\simeq 1/\eps$. By virtue of Stirling's formula, 
it is of order
$\exp(-1/\eps)/\sqrt{\eps}$ as $\eps \ra 0$. The optimal truncation method
was first used for the adiabatic approximation by Berry in 
\cite{b1}. He constructed approximate solutions to (\ref{SE})
and gave heuristic arguments concerning their exponential accuracy
and the determination of the 
exponentially small transition mechanism between asymptotic eigenstates.
In particular, the switching function (erf+1)/2 in Theorem 1.1
first appeared in \cite{b1}.
Berry further claimed that this function was universal, {\it i.e.}, 
the time development of non--adiabatic transitions in all systems
governed by one crossing point (and its conjugate) were described by
this switching function.
His formal arguments were supported by beautiful
numerical investigations of Lim and Berry \cite{lb}. 

Berry's paper \cite{b1} is the main inspiration for the present work.

\vskip .5cm
Mathematically rigorous
exponential bounds for solutions to (\ref{SE}) (and of 
${\cal A}(\eps)$) using optimal truncation were first obtained
in the general situations by Nenciu in \cite{n4}.
They were refined later in \cite{jp2}. 
See \cite{hj7} for an elementary derivation of such results. 

Although these rigorous results prove the exponential accuracy 
of the optimal truncation technique, their estimates are not 
accurate enough to capture the exponentially smaller non--adiabatic 
transitions.
An exponentially small bound on ${\cal A}(\eps)$ is an easy corollary,
but the estimates do not provide the asymptotic leading term
(\ref{trapro1}) for ${\cal A}(\eps)$.

We refer the reader to two fairly recent reviews, \cite{iamp} and \cite{ae}, 
for more details and many other aspects of the adiabatic approximation in 
quantum mechanics.
 
\vskip .5cm
We also note that
in the broader context of singular perturbations of linear ODE's, 
Theorem 1.1 can be interpreted as the smooth crossing of a Stokes 
line that emanates from some eigenvalue crossing point or singularity. 
See {\it e.g.}, \cite{b0}, 
\cite{mcleod}, \cite{daalhuis}, \cite{howls}, \cite{ck} and references 
therein. However, from our perspective, in
all rigorous work that has dealt with such issues, 
the crossing of a Stokes 
line is performed on a very small circle around the point
responsible for the transition. In this paper all estimates are performed
on the real axis, and the two conjugate 
points responsible for the transition are 
fixed and away from the real axis. 

Another angle of attack for such problems uses Borel summation
ideas. This is done for certain singularly perturbed ODE's {\it e.g.}, in 
\cite{costin}. The method consists of writing the solution
as a Laplace transform evaluated at $1/\eps$, {\it i.e.},
as $\ds \int_0^{\infty}\,F(t,\,p)\,e^{p/\eps}\,dp$.
One then derives a PDE for $F$.
The PDE is roughly what one obtains from the original equation by
replacing $1/\eps$ by the symbol $\partial/\partial p$.
To get exponentially precise information about the solution to the original
problem, it is enough to study the location
and nature of complex singularities of $F$. 

We tried to implement the Borel summation technique for our problem, but
failed to obtain sufficiently detailed information on the nature
of the singularities.

\vskip .5cm
To the best of our knowledge,
no rigorous results that address the issue
we describe in Theorem 1.1 are available in the literature.

\vskip .5cm
Before we turn to the proof of Theorem 1.1, let us briefly discuss our
choice of Hamiltonian (\ref{H}). This choice belongs to the 
family of real--symmetric time--dependent $2\times 2$ Hamiltonians
with non--degenerate eigenvalues $E_2(t)>E_1(t)$. For any member of that
family, we can assume without loss 
of generality that $E_1(t)=-E_2(t)$, because we can subtract a 
time--dependent multiple of the identity from $H(t)$,
which only changes the solutions by a trivial phase.
If we change the time variable from $t$ to
$t'\,=\,2\,\int_0^t E_2(s)\,ds/E$ and drop the prime on $t'$,
we obtain a Schr\"odinger equation (\ref{SE}) with a new Hamiltonian
$h(t)$ of the form
\be
h(t)\ =\
\frac E2 \left(\begin{array}{cc}\cos(\alpha(t))&\sin(\alpha(t))\\ 
\sin(\alpha(t))&-\cos(\alpha(t))
\end{array}\right)\,
\ee
whose eigenvalues are $\pm E/2$ for every $t$. The angle $\alpha$ is given
by some function of time.
We choose
\bea\nonumber
\Phi_1(t)=\pmatrix{-\sin(\alpha(t)/2)\cr \cos(\alpha((t)/2)}, \qquad
\mbox{and}\qquad
\Phi_2(t)=\pmatrix{\cos(\alpha(t)/2)\cr \sin(\alpha((t)/2)},
\eea
as the normalized real eigenvectors of
$h(t)$. Then the coupling $f(t)$ that drives transitions between the 
instantaneous eigenvectors is given by
\bea\nonumber
f(t)\ =\ \langle\,\Phi_2(t),\,\Phi_1'(t)\,\rangle
\ =\ -\ \langle\,\Phi_1(t),\,\Phi_2'(t)\,\rangle
\ =\ -\ \frac {\alpha'(t)}{2}.
\eea

Our choice of Hamiltonian (\ref{H}) corresponds to  
\bea\nonumber
f(t)\ =\ \frac{1}{2(t^2+\delta^2)} \qquad \Leftrightarrow \qquad 
\alpha(t)\ =\ -\ \frac 1\delta\ \arctan(t/\delta),
\eea
where $\delta>0$ is a parameter monitoring 
the strength of the coupling. 
This choice of coupling presents the 
simplest non-trivial singularities in the complex $t$-plane.

From another point of view, our choice of Hamiltonian is motivated by the
Landau--Zener Hamiltonian,
$\ds
\left(\,\begin{array}{cc}\delta&t\\t&-\delta\end{array}\,\right),
$
which has the local structure of a
generic avoided crossing \cite{gah}. Physically, one expects the transition
to take place in a neighborhood of $t=0$, and one expects that
only the form of the Hamiltonian for small $t$ should determine the
transition dynamics. To first order near $t=0$, our Hamiltonian agrees
with the Landau--Zener Hamiltonian.

However, quantum transitions have a more global character.
The presence of the square root factor in (\ref{H}),  
gives rise to the nontrivial prefactor $\sqrt{2}$ in the transition
amplitude. Indeed, it is shown in \cite{j0} that a change of  
variable allows one to transform equation (\ref{SE}) with Hamiltonian 
(\ref{H}) into an equivalent Schr\"odinger equation driven by a Hamiltonian 
that behaves locally near $t=0$ as a Landau--Zener type Hamiltonian,
but with a non-generic complex crossing point. This non-generic 
structure is responsible for the nontrivial prefactor $\sqrt{2}$ in the 
transition amplitude. It follows from 
\cite{j0}, Section 4, that the leading order transition amplitude 
for our Hamiltonian (\ref{H}) is
\be\label{oldstuff}
{\cal A(\eps)}\simeq \sqrt{2}e^{-E\,\delta /\eps}.
\ee

\vskip .5cm
The rest of the paper is organized as follows. In the next section we develop 
perturbation expansions for solutions to the time--dependent Schr\"odinger 
equation (\ref{SE}) with Hamiltonian (\ref{H}).
In Section \ref{ARR} we analyze the behavior of the high order terms
of this expansion that are required for precise optimal truncation.
In Section \ref{truncation} 
we study the error term obtained from optimal truncation and
define optimal adiabatic states.  Section \ref{integral}
is devoted to the study of two integrals that arise in the error term and
give rise to the switching function.
Theorem 1.1 is then proven in Section \ref{July4}.

\vskip .5cm
\noindent{\bf Acknowledgements}\quad  George Hagedorn wishes to thank
the Institut Fourier of the Universit\'e de Grenoble I for its
kind hospitality and support. Alain Joye wishes to thank Virgina Tech
for its kind hospitality and the NSF for travel support. We also wish to
thank Ovidiu Costin for many useful discussions about this problem.

\vskip .75cm
\section{The Formal Perturbation Expansion}\label{FPE}
\setcounter{equation}{0}

We start by converting our time--dependent
Schr\"odinger equation (\ref{SE}) with Hamiltonian 
(\ref{H}) into a parameter free equation.

Changing variables from $t$ to $s$ with $s=t/\delta$, we obtain
\bea\nonumber i\,\eps\,\frac{\partial\psi}{\partial s}\ =\ 
\frac {E\,\delta}{2\,\sqrt{1+s^2}}
\left(\begin{array}{cc}1&s\\s&-1\end{array}\right)\,\psi.
\eea
Thus, without loss of generality, by changing
$\epsilon$ into $\epsilon'=\epsilon/(E\delta)$,
we can study the parameter free model with 
$\delta=E=1$ in (\ref{SE}). We drop the prime, 
keep $\epsilon$ in the notation, and consider the coupling
\be\label{rencou}
f(t)=\frac 1{2\,(1+t^2)}.
\ee

\vskip .5cm
We now develop a formal asymptotic expansion to solutions of (\ref{SE}).

We concentrate on constructing a formal perturbation expansion
of the solution to (\ref{SE}) that corresponds to the negative
eigenvalue $-1/2$ for small $\eps$. 
We make the unusual ansatz that (\ref{SE}) has a formal
solution of the
form
\be\label{ansatz}
\psi(\eps,\,t)\ =\ e^{it/(2\eps)}\ 
e^{\int_0^t\,f(s)\,g(\eps,\,s)\,ds}\ 
\left(\,\Phi_1(t)\,+\,g(\eps,\,t)\,\Phi_2(t)\,\right),
\ee
where $\ds g(\eps,\,t)\,=\,\sum_{j=1}^{\infty}\,g_j(t)\,\eps^j$.

\vskip .5cm
\noindent{\bf Remarks}\\ 
{\bf 1.}
We arrived at this ansatz by attempting a formal solution of the form
\bea\nonumber
\psi(\eps,\,t)\ =\ e^{it/(2\eps)}\ e^{z(\eps,\,t)}\
\left(\,\Phi_1(t)\ +\ g(\eps,\,t)\ \Phi_2(t)\,\right),
\eea
with $\ds z(\eps,\,t)\,=\,\sum_{j=1}^{\infty}\,z_j(t)\,\eps^j$.
We then realized that this required
$\ds z(\eps,\,t)\,=\,\int_0^t\,f(s)\,g(\eps,\,s)\,ds$.\newline
{\bf 2.} There are more standard ans\"atze for the perturbation
expansion \cite{b1,b2,hj7,j,jkp,jp1,jp2,iamp,m,n1,n2,n3,n4,sj},
but we were unable to get sufficient control
of their $n^{\mbox{\scriptsize th}}$ terms to prove the 
estimates that we required.\newline
{\bf 3.} If we do not expand $g(\eps,\,t)$, then we find that it
must satisfy
\bea\nonumber
i\,\eps\,g'(\eps,\,t)\ =\,g(\eps,\,t)\,-\,
i\,\eps\,f(t)\,\left(\,1\,+\,g(\eps,\,t)^2\,\right).
\eea
However, we will not use this equation.
\newline
{\bf 4.} For normalization purposes, we later consider 
(\ref{ansatz}) with $\int_{-\infty}^tf(s)g(\eps,s)\,ds$ in the exponent, 
instead of $\int_{0}^tf(s)g(\eps,s)\,ds$. 
\newline
{\bf 5.} When seeking a solution that corresponds to the positive 
eigenvalue for small $\eps$, one makes the similar ansatz
\be\label{ansatz2}
\psi(\eps,\,t)\ =\ e^{-it/(2\eps)}\ 
e^{-\int_0^t\,f(s)\,\tilde{g}(\eps,\,s)\,ds}\ 
\left(\,\Phi_2(t)\,+\,\tilde{g}(\eps,\,t)\,\Phi_1(t)\,\right),
\ee
where $\ds 
\tilde{g}(\eps,\,t)\,=\,\sum_{j=1}^{\infty}\,\tilde{g}_j(t)\,\eps^j$
satisfies
\bea\nonumber
i\,\eps\,\tilde{g}'(\eps,\,t)\ =-\,\tilde{g}(\eps,\,t)\,+\,
i\,\eps\,f(t)\,\left(\,1\,+\,\tilde{g}(\eps,\,t)^2\,\right).
\eea
Hence, for any $j\in\N$, we have $\tilde{g}_j(t)=g_j(-t)$.

\vskip .5cm
We substitute (\ref{ansatz}) into (\ref{SE}) and formally solve
the resulting equation order by order in powers of $\eps$.

\noindent{\bf First Order}:\quad The terms of order $\eps$ require
\be\label{g1eqn}
g_1(t)\ =\ i\,f(t).
\ee

\noindent{\bf Second Order}:\quad The terms of order $\eps^2$ require
\be\label{g2eqn}
g_2(t)\ =\ i\,g_1'(t).
\ee

\noindent{\bf Third and Higher Order}:\quad The terms of order 
$\eps^{n+1}$
for $n\ge 2$ require
\be\label{gneqn}
g_{n+1}(t)\ =\ i\, \left(\,g_n'(t)\,+\,
f(t)\,\sum_{j=1}^{n-1}\,g_j(t)\,g_{n-j}(t)\,\right).
\ee

\vskip .5cm
Using (\ref{rencou}) for the coupling $f$, and an easy 
induction using partial fractions decompositions,
we see that $g_n(t)$ can be written as
\bea\label{decomposeg}
g_n(t)\ =\ \sum_{j=1}^{2n}\,c_{n,j}\,e_j(t),
\ \ \mbox{where } \ \ e_{2j-1}(t)\,=\,(1+it)^{-j}, \ \ 
e_{2j}(t)\,=\,(1-it)^{-j}.
\eea
Thus, for each $n$, we can associate $g_n$ with a unique element
of $l^1$, the space of absolutely summable sequences.

Following the intuition of Michael Berry \cite{b1,b2}, we isolate the
highest order poles of $g_n(t)$ by decomposing
$g_n(t)\,=\,G_n(t)\,+\,h_n(t)$,
where $G_n(t)\,=\,c_{n,2n-1}\,e_{2n-1}(t)\,+\,c_{n,2n}\,e_{2n}(t)$.

Proof of our results now depends on an analysis of the behavior of
$G_n(t)$ and $h_n(t)$ for large $n$.

\vskip .75cm
\section{Analysis of the Recurrence Relation}\label{ARR}
\setcounter{equation}{0}

\vskip .5cm
The main goals of this section are summarized in the
following proposition.

\vskip .5cm
\begin{prop}\label{gbounds} There exists $C$, such that
for all real $t$ and each $n\ge 1$, 
\bea\label{Gbound}
|G_n(t)|&\le&\,(n-1)!,\\[5pt]
\label{DGbound}
|G_n'(t)|&\le&\,n!,\\[5pt]
\label{hbound}
|h_n(t)|&\le&C\,(n-2)!\,\log(n-2),\\[5pt]
\label{Dhbound}
|h_n'(t)|&\le&C\,(n-1)!\,\log(n-2).
\eea
\end{prop}

\vskip .5cm
We prove this result using the recurrence formulas 
(\ref{g1eqn})--(\ref{gneqn}) and the following amazing fact
that we use to control the nonlinear terms in the recurrence
relation.

\vskip .5cm
\begin{lem}\label{amazing} For each $k$ and $m$,
$\ds e_k(t)\,e_m(t)\ =\ \sum_{j=1}^{k+m+1}\,d_{k,m,j}\,e_j(t)$.
\newline Every $d_{k,m,j}$ is a non-negative real number, and
$\ds \sum_{j=1}^{k+m+1}\,d_{k,m,j}\ =\ 1$.
\end{lem}

\vskip .5cm
\noindent{\bf Remark}\quad This lemma implies that we have a
Banach algebra structure on $l^1$, where the product of
$\{a_n\}$ and $\{b_n\}$ is determined by formally multiplying
$\sum_n\,a_n\,e_n(t)$ times $\sum_n\,b_n\,e_n(t)$ and then taking
the coordinates of the result in the $\{e_j(t)\}$ basis. 

\vskip .5cm
\noindent{\bf Proof of Lemma \ref{amazing}}\quad
We first remark that by keeping track of the orders of the poles, 
it is easy to see that $j\le k+m+1$ in the sums in the lemma.

Next, we observe that there are many trivial situations.
If $k$ and $m$ are both odd, then $e_k(t)\,e_m(t)=e_{k+m+1}(t)$.
If $k$ and $m$ are both even, then $e_k(t)\,e_m(t)=e_{k+m}(t)$.

Thus, the only non-trivial cases are when one is odd and the other
is even. 
To prove these cases, we do inductions on odd $k$ and even $m$.

For $k=1$ and $m=2$, the lemma follows immediately from 
\be\label{pf}
(1+it)^{-1}\ (1-it)^{-1}\ =\ 
\frac 1{2}\,(1+it)^{-1}\ +\ \frac 1{2}\,(1-it)^{-1}.
\ee

Next, we fix $m=2$ and assume inductively that 
the lemma has been proven for $k=2K-1$. 
Then using (\ref{pf}) again, we have
\bea\nonumber
e_{k+2}(t)\,e_m(t)&=&(1+it)^{-K-1}\,(1-it)^{-1}\\[5pt]
\nonumber
&=&(1+it)^{-K}\,\left[\,\frac 1{2}\,(1+it)^{-1}\,+\,
\frac 1{2}\,(1-it)^{-1}\,\right]\\[5pt]
&=&\frac 1{2}\,\left[\,e_{k+2}(t)\,+e_{k}(t)e_2(t)\,\right]
\eea
The result now follows from our induction hypothesis.

Thus, the lemma is true for all $k$ and $m=2$.

Finally, we fix an odd $k$, and assume inductively that
the lemma has been proven for this $k$ and an even $m$. Then
\bea\nonumber
e_{k}(t)\,e_{m+2}(t)&=&(e_k(t)\,e_m(t))\,e_2(t)\\[5pt]
\nonumber
&=&\sum_j\,d_{k,m,j}\,e_j(t)\,e_2(t)\\[5pt]
\nonumber
&=&\sum_{j}\,d_{k,m,j}\,\sum_{j'}\,d_{j,2,j'}\,e_{j'}(t).
\eea
The lemma now follows since $\sum_{j'}\,d_{j,2,j'}\,=\,1,
\quad\sum_j\,d_{k,m,j}\,=\,1$, and all the 
$d$'s that occur here are non-negative.\ep

\vskip .5cm
We also need the following technical result.

\vskip .5cm
\begin{lem}\label{silly} For each $n\ge 1$, we have
\bea\label{silly1}
\sum_{j=0}^n\,\frac{(n-j)!\ j!}{n!}&\le&\frac 83,\\[5pt]
\nonumber
\sum_{j=0}^{n-1}\,\frac{(n-j)!\ j!}{n!}&\le&\frac 53,
\quad\mbox{and}\\[5pt]
\nonumber
\sum_{j=1}^{n-1}\,\frac{(n-j)!\ j!}{n!}&\le&\frac 23.
\eea
\end{lem}

\vskip .5cm
\noindent{\bf Proof}\quad 
The first inequality trivially implies the other two.

We observe by direct computation
that the result is true for the first few values of $n$,
and that the sum equals $8/3$ when $n=3$ and $n=4$. 

For $n\ge 5$, we separate the first two terms and 
last two terms to see that the sum equals
\bea\nonumber
2\,+\,\frac 2n\,+\,\sum_{j=2}^{n-2}\,\frac{(n-j)!\ j!}{n!}.
\eea
The largest terms in the sum over $j$ in this expression come
from $j=2$ and $j=n-2$. Those terms equal
$\ds \frac 2{n(n-1)}$, and there are $(n-3)$ terms. Thus, the
left hand side of (\ref{silly1}) is bounded by
\bea\nonumber
2\,+\,\frac 2n\,+\,\frac{2(n-3)}{n(n-1)}
\ \le\ 2\,+\,\frac 25\,+\,\frac 15
\ <\ \frac 83.
\eea
This last step relies on the observation that
$\frac{2(n-3)}{n(n-1)}$ takes the value $1/5$
when $n=5$ and $n=6$, and that it is decreasing for 
$n\ge 6$.\ep

\vskip .5cm
For any $y(t)\,=\,\sum_j\,y_j\,e_j(t)$, with $\{y_j\}\in l^1$,
we define $\|y\|\,=\,\sum_j\,|y_j|$. We note that for $t\in \R$,
$|y(t)|\,\le\,\|y\|$. Since $G_n$ is obtained from $g_n$ by
dropping components in the $e_j(t)$ basis, we note that
$\|G_n\|\,\le\,\|g_n\|$.
Thus, the following lemma implies (\ref{Gbound}) and
(\ref{DGbound}) since
$\frac{d\phantom{t}}{dt}\,(1\pm it)^{-j}\,=\,
\mp\,i\,j\,(1\pm it)^{-j-1}$.

\vskip .5cm
\begin{lem}\label{gbound}$\|g_n\|\,\le\,(n-1)!$.
\end{lem}

\vskip .5cm
\noindent{\bf Proof}\quad
We prove that the sequence 
$a_n\,=\,\|g_n\|/(n-1)!$ is bounded above by $1$.

By (\ref{gneqn}), Lemmas \ref{amazing} and \ref{silly},
we see that $n\,\ge\,2$ implies
\be\label{arecur}
a_{n+1}\ \le\ a_n\,+\,\frac{4\,a_{n-1}^2}{3n(n-1)}.
\ee
From (\ref{g1eqn}) and (\ref{g2eqn}) we have
$a_1\,=\,a_2\,=\,1/2$. By explicit computation, we observe
that
$a_{3}=\frac{17}{32}\,\le\,1-\frac {4}{9}$, and 
$a_{4}=\frac{197}{384}\,\le\,1-\frac 4{12}$.

The lemma now follows by induction (starting at $n=4$)
and the following statement:\quad
If $n\ge 4,\ $ 
$a_{n-1}\,\le\,1\,-\,\frac{4}{3(n-1)}$
and 
$a_n\,\le\,1\,-\,\frac{4}{3n}$, then
$a_{n+1}\,\le\,1\,-\,\frac{4}{3(n+1)}$.

To prove this statement, we use (\ref{arecur}) to see that
for $n\geq 4$
\bea\nonumber
a_{n+1}&\le&1\,-\,\frac{4}{3n}\,+\,
\frac{4}{3n(n-1)}\,\left(\,1-\frac 4{3(n-1)}\,\right)^2\\[5pt]
\nonumber
&=&1\,-\,\frac{4}{3(n+1)}\,-\,
\frac{8(3n^2+10n-29)}{27(n+1)n(n-1)^3}\\[5pt]
\nonumber
&\leq&1\,-\,\frac{4}{3(n+1)}.\ep
\eea

\vskip .5cm
Lemma \ref{gbounds} is now proven by the comments
before Lemma \ref{gbound} and the following lemma. 

\vskip .5cm
\begin{lem}\label{hbound1} There exists $C$, such that
$\|h_n\|\,\le\,C\,(n-2)!\,\log(n-2)$
and \newline
$\|h_n'\|\,\le\,C\,(n-1)!\,\log(n-2)$.
\end{lem}

\vskip .5cm

\noindent{\bf Proof}\quad The first estimate implies the
second since $h_n$ is in the span of $e_j(t)$ with
$j\le 2n-2$.

Define $b_n\,=\,\|h_n\|/((n-2)!)$.
Since $h_n$ is obtained from $g_n$ by dropping components
in the $e_j(t)$ basis, we have $\|h_n\|\,\le\,\|g_n\|$.
Thus, by Lemma \ref{gbound}, $b_n\,\le\,n-1$.

We rewrite (\ref{gneqn}), using $g_n\,=\,G_n\,+\,h_n$:
\bea\nonumber\hspace{-29pt}
G_{n+1}\,+\,h_{n+1}\ =\ i\,\left(
\,\frac{dG_n}{dt}\,+\,\frac{dh_n}{dt}\,+\,
f\,\sum_{j=1}^{n-1}\,G_j\,G_{n-j}\,+\,
2\,f\,\sum_{j=1}^{n-1}\,G_j\,h_{n-j}\,+\,
f\,\sum_{j=1}^{n-1}\,h_j\,h_{n-j}\,\right).
\eea
We then drop the $e_{2n+1}(t)$ and $e_{2n+2}(t)$ components 
of this expression to obtain an expression for $h_{n+1}$.
This involves dropping the entire term $\ds i\,\frac{dG_n}{dt}$,
as well as parts of other terms. Since the norm decreases when
we drop components, we see that
\bea\nonumber\hspace{-15pt}
\|h_{n+1}\|\ \le\ \left(\,
\left\|\frac{dh_n}{dt}\right\|\,+\,
\left\|f\,\sum_{j=1}^{n-1}\,G_j\,G_{n-j}\right\|\,+\,
2\,\left\|f\,\sum_{j=1}^{n-1}\,G_j\,h_{n-j}\right\|\,+\,
\left\|f\,\sum_{j=1}^{n-1}\,h_j\,h_{n-j}\right\|\,\right).
\eea
Since $h_n$ is in the span of $\{e_j(t)\}$ for $j\le 2n-2$,
$\ds \left\|\frac{dh_n}{dt}\right\|\,\le\,(n-1)\,\|h_n\|$.
Thus, by $\|f\|\,\le\,1/2$, $\|G_n\|\,\le\,(n-1)!$,
$h_1\,=\,h_2\,=\,0$, Lemmas \ref{amazing}, and \ref{silly}, we have
\bea\nonumber
b_{n+1}\ \le\ b_n\,+\,\frac{4/3}{n-1}\,+\,
\frac{5/3}{(n-1)(n-2)}\ b_{n-1}\,+\,
\frac{1/3}{(n-1)(n-2)(n-3)}\ b_{n-3}^2.
\eea
Since we already have $b_n\,\le\,n-1$ and
$b_1\,=\,b_2\,=\,0$, this implies
\bea\nonumber
b_{n+1}\ \le\ b_n\,+\,\frac{4/3}{n-1}\,+\,
\frac{5/3}{n-1}\,+\,\frac{1/3}{n-1}
\ =\ b_n+\frac{10/3}{(n-1)},
\eea
for $n\ge 2$.
Thus,
\bea\nonumber
b_n\ =\
\sum_{k=2}^{n-1}\,\left(\,b_{k+1}\,-\,b_k\,\right)
\ =\ \sum_{k=2}^{n-1}\,\frac{10/3}{k-1}
\ \le\ \frac{10}3\,\left(\,1\,+\,\log(n-2)\,\right).
\eea
This implies the lemma and completes the proof of Proposition 
\ref{gbounds}.\ep

\vskip .5cm
The functions $e_j(t)$ are in $L^1(\R)$ for $j\ge 3$.
The functions $e_1(t)$ and $e_2(t)$ are not in $L^1(\R)$, 
but $e_1(t)+e_2(t)$ is. The following lemma facilitates getting
$L^1$ information about the functions $g_n$.

\vskip .5cm
\begin{lem}\label{nutty}
For every $n$, the $e_1(t)$ and $e_2(t)$ coefficients
in $g_n(t)$ are equal.
Furthermore, the $L^1(\R)$ norm of
$g_n$ is bounded by $\pi\ \|g_n\|$.
These results are also true for $h_n$.
\end{lem}

\vskip .5cm
\noindent{\bf Proof}\quad
We prove the first statement by induction on $n$.
It is clearly true for $n=1$.
Suppose it is true for all $n<N$. The function
$i\,g_{N-1}'$ contains no $e_1(t)$ or $e_2(t)$ component.
Since the $g_n$ are bounded,
$f\,\sum_{j=1}^{N-2}\,g_j\,g_{N-j-1}$ is in $L^1(\R)$ since $f$
is. The only way this function can be in $L^1$ is if it's $e_1(t)$
and $e_2(t)$ components are equal. This implies the result for
$g_N$, and the induction can proceed.

The second statement follows from the first because
the absolute values of $(e_1(t)+e_2(t))$ and $e_j(t)$ for $j\ge 3$
are dominated by $(1+t^2)^{-1}$ which has integral $\pi$.

The third statement follows since $h_n$ is obtained from $g_n$
by removal of the $n^{\mbox{\scriptsize th}}$ order pole terms.
\ep

\vskip .5cm
We now examine $G_n$ more closely. We note that highest order pole
terms in $g_n$ at $t=\pm i$ satisfy the recurrence relation
\bea\nonumber
G^{\pm}_{n+1}(t)\ =\ 
\left(\,\,i\,\frac{dG_n^{\pm}}{dt}(t)\ \pm\ 
\frac{1/4}{t\mp i}\ 
\sum_{j=1}^{n-1}\ G_j^{\pm}(t)\ G_{n-j}^{\pm}(t)\,\right),
\eea
with $\ds G^{\pm}_1(t)\ =\pm\ \frac{1}{4}\ \frac 1{t\mp i}$ and
$\ds G^{\pm}_2(t)\ =\ \frac{\mp\,i}{4}\ \frac 1{(t\mp i)^2}$.
From this it follows that
\be\label{Gexpression}
G_n(t)\ =\ i\,\gamma_n\,
\left(\,e_{2n-1}(t)\,+\,(-1)^{n-1}\,e_{2n}(t)\,\right),
\ee
where $\gamma_n$ satisfies the real numerical recurrence relation
\be\label{gamma}
\gamma_{n+1}\ =\ n\ \gamma_n\ -\
\frac 14\ \sum_{j=1}^{n-1}\ \gamma_j\ \gamma_{n-j},
\ee 
with $\gamma_1\,=\,\gamma_2\,=\,1/4$.

By Lemma \ref{gbounds}, the quantity
$\beta_n\,=\,\gamma_n/(n-1)!$ is bounded. It satisfies
\be\label{beta}
\beta_{n+1}\ =\ \beta_n\ -\
\frac 14\ \sum_{j=1}^{n-1}\ 
\frac{((j-1)!)((n-j-1)!)}{n!}\ \beta_j\ \beta_{n-j},
\ee 
with $\beta_1\,=\,\beta_2\,=\,1/4$. From this relation
and Lemma \ref{silly}, it follows that $\beta_n$ has a limit
$\beta^*$ as $n$ tends to infinity. To see this, suppose
that the sequence $\beta_m$ is positive and 
strictly decreasing for $m\le n$. Then,
\be
0\ <\ \sum_{j=1}^{n-1}\ 
\frac{((j-1)!)((n-j-1)!)}{n!}\ \beta_j\ \beta_{n-j}\ <\ 
\frac{8}{3}\ \frac{\beta_1^2}{n(n-1)}.
\ee
Thus, iterating and using $\beta_1=\beta_2=1/4$, we have
\bea
\beta_{n+1}&>& \beta_n\ -\ \frac{2}{3}\ \frac{\beta_1^2}{n(n-1)}
\ >\ \beta_2\ -\ 
\frac{2\beta_1^2}{3}\ \sum_{j=2}^n\ \frac{1}{j(j-1)}\nonumber\\
&=&\beta_2\ -\ 
\frac{2\beta_1^2}{3}\ \left(1-\frac{1}{n}\right)
\ >\ \frac 5{24}.
\eea
Therefore, the sequence $\{\beta_n\}$ is positive, strictly decreasing and
bounded below.

Similarly, for some constant $C>0$ and any $p>0$, we have
\be
\beta_{n}\ -\ \beta_{n+p}\ \leq\ 
C\ \left(\,\frac1{n-1}\,-\,\frac 1{n+p-1}\,\right),
\ee
so that,
\be\label{stupid}
\beta_n\ =\ \beta^*\ (1\ +\ O(n^{-1})).
\ee

\vskip .5cm
\noindent{\bf Remark}\quad
Later, we will see that $\ds \beta^*=\frac{1}{\pi \sqrt{2}}$.

\section{Optimal Truncation}\label{truncation}
\setcounter{equation}{0}

We begin this section by studying
\be\label{zeta}
\zeta_n(\eps,\,t)\ = \
i\,\eps\,\frac{\partial\psi}{\partial t}\,-\,H(t)\,\psi,
\ee
where $\psi$ is given by (\ref{ansatz}) with
$\ds g(\eps,\,t)\,=\,\sum_{j=1}^{n}\,g_j(t)\,\eps^j$.
We ultimately choose $n=\grintl 1/\eps\grintr-1$,
where $\grintl k\grintr$ denotes the greatest integer
less than or equal to $k$.

By explicit calculation, $\zeta_n(\eps,\,t)$ equals 
$e^{it/(2\eps)}\ e^{\int_0^t\,(f(s)\,
\sum_{j=1}^{n}\,g_j(s)\,\eps^j)\,ds}\ \Phi_2(t)$ times
\be\label{zeta1}
i\,\eps^{n+1}\,G_n'\,+\,i\,\eps^{n+1}\,h_n'\ +\
i\,\eps^{n+1}\,f\,\sum_{j=1}^{n-1}\,g_j\,g_{n-j}
\ +\ \sum_{k=n+2}^{2n+1}\,
i\,\eps^{k}\,f\sum_{j=k-n-1}^n\,g_j\,g_{k-j-1}.
\ee

\vskip .5cm
\begin{lem}\label{yeah}
The first term in (\ref{zeta1}) satisfies
\be\label{berryterm}
\eps^{n+1}\,\|G_n'\|\ =\ 2\,\beta_n\,\eps^{n+1}\,(n!).
\ee
When $n$ is
chosen to be $n\,=\,\grintl 1/\eps\grintr\,-\,1$, 
the norm of the remaining terms in (\ref{zeta1}) satisfies
\be\label{sneak}
\left\|\,\zeta_n(\eps,\,t)\,-\,i\,\eps^{n+1}\,G_n'\,\right\|
\ \le\ 
C\,\eps^{n+1}\,((n-1)!)\,\log(n-2).
\ee
 for some $C$. Thus,
as $\eps$ tends to zero with
$n\,=\,\grintl 1/\eps\grintr\,-\,1$,\quad
$\zeta_n(\eps,\,t)$ is asymptotic to $i\,\eps^{n+1}\,G_n'(t)\
e^{ it/(2\eps)}\ e^{\int_0^t\,(f(s)\,
\sum_{j=1}^{n}\,g_j(s)\,\eps^j)\,ds}\ \Phi_2(t)$.
\end{lem}

\vskip .5cm
\noindent{\bf Proof}\quad
The result (\ref{berryterm}) was proven at the end of the
previous section.

By Lemma \ref{hbound1}, the second term in (\ref{zeta1}) satisfies
\be\label{secondterm}
\eps^{n+1}\,\|h_n'\|\ \le\ C'\,\eps^{n+1}\,(n-1)!\,\log(n-2).
\ee
By Lemmas \ref{amazing}, \ref{silly}, and \ref{gbound}, the third
term satisfies
\be\label{thirdterm}
\eps^{n+1}\,\left\|\,f\,\sum_{j=1}^{n-1}\,g_j\,g_{n-j}\,\right\|
\ \le\ \eps^{n+1}\ \frac 83\ \frac 12\ (n-2)!
\ =\ \frac 43\ \eps^{n+1}\ (n-2)!.
\ee
We now prove that
\be\label{finally}
\left\|\,\sum_{k=n+2}^{2n+1}\,
\eps^{k}\,f\sum_{j=k-n-1}^n\,g_j\,g_{k-j-1}\,\right\|
\ \le\ C''\,\eps^{n+1}\,(n-1)!
\ee
for some $C''$ when $n\,=\,\grintl 1/\eps\grintr\,-\,1$.

We begin the proof of (\ref{finally}) with a technical lemma:

\vskip .5cm
\begin{lem}\label{silly2} For positive integers $l\le m/2$,\quad
$\ds \sum_{p=l}^{m-l}\ (p!)\,((m-p)!)\ \le\ \,(m-2l+1)\,(l!)\,((m-l)!)$.
\end{lem}

\noindent{\bf Proof}\quad 
\bea\nonumber
&&\sum_{p=l}^{m-l}\ (p!)\,((m-p)!)\\[5pt]
\nonumber &=&(l!)((m-l)!)\,+\,((l+1)!)((m-l-1)!)\,+\,\cdots \\[5pt]
\nonumber &&+\,((m-l-1)!)((l+1)!)\,+\,((m-l)!)(l!).
\eea
The first and last terms are the largest terms in this sum, and they
equal $(l!)\,((m-l)!)$. The lemma follows
since there are $(m-2l+1)$ terms in the sum.\ep

\vskip .5cm
We apply Lemma \ref{silly2} with $p=j-1$, $m=k-3$, and $l=k-n-2$, along
with Lemmas \ref{amazing}, \ref{silly}, and \ref{gbound}, to see that 
the norm of the term
$\ds i\,\eps^{k}\,f\sum_{j=k-n-1}^n\,g_j\,g_{k-j-1}$ in (\ref{zeta1}) is
bounded by 
$\ds \eps^k\,\frac 12\,\sum_{p=l}^{m-l}\,p!\,((m-p)!)
\ \le\ \eps^k\,(2n-k+2)\,((k-n-2)!)\,((n-1)!)/2$.

To prove (\ref{finally}), we now sum this quantity over $k$:
\bea\nonumber
&&\sum_{k=n+2}^{2n+1}\,\eps^k\,(2n-k+2)\,((k-n-2)!)\,((n-1)!)/2
\\[5pt]\nonumber
&=&\eps^{n+1}\,((n-1)!)\,
\sum_{k=n+2}^{2n+1}\,\frac{2n-k+2}{2}\,\eps^{k-n-1}\,
((k-n-2)!)\\[5pt]\label{tired}
&=&\eps^{n+1}\,((n-1)!)\,
\sum_{m=0}^{n-1}\,\frac{n-m}{2}\,\eps^{m+1}\,(m!).
\eea

We now fix $n\,=\,\grintl 1/\eps\grintr\,-\,1$
and note that this implies
that
\bea\nonumber
&&\sum_{m=0}^{n-1}\,\frac{n-m}{2}\,\eps^{m+1}\,(m!)\\[5pt]
\nonumber
&\le&\frac 12\,\sum_{m=0}^{n-1}\,\,\eps^{m}\,(m!)\\[5pt]
\nonumber
&\le&\frac 12\,\left(\,1\,+\,\frac 1n\,+\,\frac {2!}{n^2}\,+\,\cdots\,
\,+\,\frac{(n-1)!}{n^{n-1}}\,\right).
\eea
By Stirling's formula, $j!/n^j\,\leq\,C\,(j/n)^j\,e^{-j}\,\sqrt{j}$
for some $C>0$. So,
the quantity on the right hand side here is bounded.
This and (\ref{tired}) imply (\ref{finally}),
which proves the lemma.\ep

\vskip .5cm
We note that our choice of $n=\grintl 1/\eps \grintr -1$ 
implies that there exists
a $C>0$ such that 
\be\label{intsum}
\left|\,\int\,\sum_{j=1}^{n}\,g_j(t)\,\eps^j\,dt\,\right|
\ \leq\ C\,\eps.
\ee
This follows from Lemmas \ref{gbound} and \ref{nutty}, since the 
left hand side of (\ref{intsum}) is bounded by
\bea\nonumber
\pi\,\sum_{j=1}^{n}\,\|g_j\|\,\eps^j\ \leq\
\pi\,\sum_{j=1}^{n}\,(j-1)!\,\eps^j\ \leq\ 
\pi\,\eps\,\sum_{k=0}^{n-1}\,\eps^k\,k!\ \leq\ C\,\eps .
\eea

\vskip .5cm
This allows us to define the optimal adiabatic state $\psi_1(\eps,t)$ 
associated with the eigenvalue $-1/2$ by
\be\label{opti1}
\psi_1(\eps,\,t)\ =\ e^{it/(2\eps)}\ 
e^{\int_{-\infty}^t\,f(s)\,g(\eps,\,s)\,ds}\ 
\left(\,\Phi_1(t)\,+\,g(\eps,\,t)\,\Phi_2(t)\,\right),
\ee
where 
$\ds g(\eps,\,t)\,=\,\sum_{j=1}^{\grintl 1/\eps \grintr-1}
\,g_j(t)\,\eps^j$,
with the $g_j(t)$'s defined in Section 2.
By construction, $\psi_1(\eps,t)$ 
is normalized as $t\ra-\infty$.
The optimal adiabatic state $\psi_2(\eps,t)$ 
associated with the eigenvalue $1/2$ is defined similarly,
according to (\ref{ansatz2}),
\be\label{opti2}
\psi_2(\eps,\,t)\ =\ e^{-it/(2\eps)}\ 
e^{-\int_{-\infty}^t\,f(s)\,\tilde{g}(\eps,\,s)\,ds}\ 
\left(\,\Phi_2(t)\,+\,\tilde{g}(\eps,\,t)\,\Phi_1(t)\,\right),
\ee
where $\ds \tilde{g}(\eps,\,t)\,=
\,\sum_{j=1}^{\grintl 1/\eps \grintr-1}\,\tilde{g}_j(t)\,\eps^j$.
Since the entire analysis of the $g_j$'s above does not depend on the 
sign of $t$, it holds for the 
$\tilde{g}_j(t)$ as well. See Remark 5 of Section 2.

\section{Analysis of Some Integrals}\label{integral}
\setcounter{equation}{0}

\vskip .5cm
The main goal of this section
is to analyze the two (quite different) integrals
\bea\nonumber
\int_{-\infty}^t\,\frac{e^{is/\eps}}{(1\pm is)^m}\,ds,
\eea
where $\eps>0$ is a small parameter,
and $m=\grintl 1/\eps \grintr$. 
By taking conjugates, we obtain the analogous results for
\bea\nonumber
\int_{-\infty}^t\,\frac{e^{-is/\eps}}{(1\mp is)^m}\,ds.
\eea

\vskip .5cm
\noindent{\bf Remark}
As we see below, the reason
the two integrals have different behavior is that in one
case, there is a cancellation of rapidly oscillating phases,
while in the other, the phases reinforce one another.

\vskip .5cm
\begin{lem}\label{favorite}
For small $\eps>0$,
$m=\grintl 1/\eps\grintr$ and any $\gamma\,\in\,(1/2,\,1)$, we have
\be\label{answer1}
\int_{-\infty}^t\,\frac{e^{is/\eps}}{(1+is)^m}\,ds
\ =\ \sqrt{\frac{\pi}{2\,m}}\,
\left\{\,\mbox{\rm erf}\left(\,
\sqrt{\frac m2}\ t\,\right)\,+\,1\,\right\}
\ +\ O(m^{-\gamma}),
\ee
and
\be\label{answer2}
\int_{-\infty}^t\,\frac{e^{is/\eps}}{(1-is)^m}\,ds
\ =\ O(m^{-\gamma}).
\ee
\end{lem}

\noindent{\bf Proof}\quad
We begin with some preliminary estimates that apply to both integrals.
We note that for real $s$,
\bea\nonumber
\left|\,\frac{e^{is/\eps}}{(1\mp is)^m}\,\right|\ =\ 
(1+s^2)^{-m/2}.
\eea
When $|s|\ge 1$, this is bounded by $|s|^{-m}$. 
Since $\ds \int_1^{\infty}\,ds/s^m\,=\,1/(m-1)$, we make an 
$O(m^{-1})$ error in each
of the integrals, if we drop the contributions from $|s|\ge 1$.

Next, let $\delta>0$ be small and $a=m^{\delta}/\sqrt{m}$. If $a\le |s|\le 1$, 
then 
$(1+s^2)^{-m/2}\,\le\,(1+a^2)^{-m/2}$.
Thus
\be\label{woof}
\int_a^1\,(1+s^2)^{-m/2}\,ds\ \le\ 
(1-a)\,(1+a^2)^{-m/2}\ \le\
(1+a^2)^{-m/2}.
\ee
Now, if $\delta$ is small enough,  (\ref{woof}) is of the order of 
\be
 e^{-\frac m2 \ln(1+a^2)}=e^{-\frac{m^{2\delta}}{2}+O(1/m^{1-4\delta})}=
O(e^{-\frac{m^{2\delta}}{2}})<< O(m^{-1}).
\ee

Therefore, if we let $A_m=\{s:\,|s|\le a=m^{\delta}/\sqrt{m}\}$, then
\be\label{reduced}
\int_{-\infty}^t\,\frac{e^{is/\eps}}{(1\pm is)^m}\,ds\ =\ 
\int_{-a}^t\,\frac{e^{is/\eps}}{(1\pm is)^m}\
\chi_{A_m}(s)\ds\ +\ O(m^{-1}).
\ee

We now concentrate on (\ref{answer1}).
\bea\nonumber
(1+is)^{-m}\ =\ (1+s^2)^{-m/2}\
\left(\,\frac{1-is}{1+is}\,\right)^{m/2}.
\eea
We separately examine the logs of the factors on the right
hand side of this equation.

First,
\bea\nonumber
\log\left((1+s^2)^{-m/2}\right)\ =\ 
-\frac m2\ \log(1+s^2).
\eea
On the support of $\chi_{A_m}$, this equals
\bea\nonumber
-\frac m2\ \left(\,s^2\,-\,s^4/2\,+\,\dots\,\right)
\ =\ -\,m\,s^2/2\,+\,O(m^{-(1-4\delta)}).
\eea
Exponentiating, we have
\bea\label{abs}
(1+s^2)^{-m/2}
\ =\ e^{-\,m\,s^2/2}\ +\ O(m^{-(1-4\delta)})\quad
\mbox{on the support of $\chi_{A_m}$.}
\eea

Second,
\bea\nonumber
\log\left(\,\left(\,\frac{1-is}{1+is}\,\right)^{m/2}\,\right)
\ =\ 
-\ i\ m\ \arctan(s)=-\ i\ m\ (s+O(s^3)).
\eea
On the support of $\chi_{A_m}$,  this equals
\bea\nonumber
-\ i\ m\ s\ +\ O(m^{-(1/2-3\delta)}).
\eea
Exponentiating, we have
\bea\label{phase}
\left(\,\frac{1-is}{1+is}\,\right)^{m/2}\ =\
e^{-\,i\,m\,s}\ +\ O(m^{-(1/2-3\delta)})\quad
\mbox{on the support of $\chi_{A_m}$.}
\eea

For $m=\grintl 1/\eps\grintr$, we have 
\be\label{exp}
e^{is/\eps}\ =\ e^{i\,m\,s}\ e^{is(1/\eps-m)}
\ =\ e^{i\,m\,s}\ \left(\,1\,+\,O(m^{-(1/2-\delta)})\,\right)
\ =\ e^{i\,m\,s}\ +\ O(m^{-(1/2-\delta)}),
\ee 
on the support of $\chi_{A_m}$.

The support of the integrand in (\ref{reduced})
has length $O(m^{-(1/2-\delta)})$.
Using this, (\ref{abs}), (\ref{phase}), and (\ref{exp})
in (\ref{reduced}) we see that for $\delta$ small enough,
\bea\nonumber
\int_{-\infty}^t\,\frac{e^{is/\eps}}{(1+is)^m}\,ds&=& 
\int_{-a}^t\,\frac{e^{is/\eps}}{(1+is)^m}\
\chi_{A_m}(s)\,ds\ +\ O(m^{-1})\\[5pt]
\nonumber
&=&\int_{-m^{-(1/2-\delta)}}^t\,e^{-\,m\,s^2/2}\
\chi_{A_m}(s)\,ds\ +\ O(m^{-(1-4\delta)}).
\eea
By simple estimates on the tail of the Gaussian,
this equals
\bea\nonumber\int_{-\infty}^t\,e^{-\,m\,s^2/2}\,ds\ +\ O(m^{-(1-4\delta)}),
\eea
which implies (\ref{answer1}).

We now turn to (\ref{answer2}). The analysis is very
similar, except that we use the conjugate of (\ref{phase}).
This leads us to
\bea\nonumber
\int_{-\infty}^t\,\frac{e^{is/\eps}}{(1-is)^m}\,ds&=& 
\int_{-a}^t\,\frac{e^{is/\eps}}{(1-is)^m}\
\chi_{A_m}(s)\,ds\ +\ O(m^{-1})\\[5pt]
\label{bark}
&=&\int_{-m^{-(1/2-\delta)}}^t\,e^{-\,m\,s^2/2}
\ e^{2\,i\,m\,s}
\chi_{A_m}(s)\,ds\ +\ O(m^{-(1-4\delta)}).
\eea
Integrating by parts, we have
\bea\nonumber
\int\,e^{-\,m\,s^2/2}\ e^{2\,i\,m\,s}\,ds\ =\
\frac 1{2\,i\,m}\ e^{-\,m\,s^2/2}\ e^{2\,i\,m\,s}\
+\ \frac 1{2i}\
\int\,s\,e^{-\,m\,s^2/2}\ e^{2\,i\,m\,s}\,ds.
\eea
Since $\ds \int_{-\infty}^{\infty}\,|s|\,
e^{-\,m\,s^2/2}\,ds\,=\,\frac 2m$,
the integral on the right hand side of 
(\ref{bark}) is
$O(1/m)$ for all $t$, and (\ref{answer2})
follows for $\delta$ small enough.\ep

\vskip .5cm
In the following section, we will also need the
following trivial estimate for any $m\ge 2$:
\bea\label{absintegral}
\int_{-\infty}^t\,
\left|\,\frac{1}{(1\pm is)^m}\,\right|\,ds
\ \le\ C.
\eea

\vskip .5cm
\section{Proof of the Theorem}\label{July4}
\setcounter{equation}{0}

\vskip .5cm
The perturbation expansions of Section \ref{FPE} generate
formal approximate solutions to (\ref{SE})
associated with the energy levels $\mp 1/2$. When
truncated at
$n=\grintl 1/\eps\grintr\,-\,1$,  the expansions 
define the optimal adiabatic states
$\psi_1(\eps,\,t)$ and $\psi_2(\eps,\,t)$ by (\ref{opti1}) and 
(\ref{opti2}). 

As we see below, each of these optimal adiabatic states
agrees with an exact solution up to an  exponentially small
error. Since $\psi_1(\eps,\,t)$ and $\psi_2(\eps,\,t)$
are asymptotically normalized for $t\ra -\infty$,
it follows that
they are normalized up to the same exponentially small error for
all $t$. Similarly, the inner product
$\langle\,\psi_1(\eps,\,t),
\,\psi_2(\eps,\,t)\,\rangle$ is exponentially small
for all $t$.

We compute the asymptotic leading term to 
the solution of (\ref{SE}) that coincides 
with $\psi_1(\eps,\,t)$ as $t \ra-\infty$,
using the optimal adiabatic states 
$\psi_1(\eps,\,t)$ and $\psi_2(\eps,\,t)$ as a basis.
This approximation is accurate up to an error of order 
$e^{-1/\eps}\eps^{\mu}$, for $0<\mu< 1/2$, uniformly
for $t\in\R$.

\vskip .5cm
We now define $\zeta(\eps,\,t)$ by (\ref{zeta}) with 
$n=\grintl 1/\eps\grintr\,-\,1$.
We denote the unitary
propagator associated with the Schr\"odinger equation
(\ref{SE}) by $U_{\eps}(t_1,\,t_2)$.
There is an exact solution $\Psi_1(\eps,\,t)$
to (\ref{SE}) that is asymptotic to $\psi_1(\eps,\,t)$
as $t\longrightarrow -\infty$. We compute
\bea\nonumber
\Psi_1(\eps,\,t)\,-\,\psi_1(\eps,\,t)
\ =\ \lim_{r\to -\infty}\,
U_{\eps}(t,\,r)\,\psi_1(\eps,\,r)\,-\,\psi_1(\eps,\,t).
\eea
In this expression, we replace
$U_{\eps}(t,\,r)\,\psi_1(\eps,\,r)$
with the integral of its derivative
with respect to $r$ (with the proper constant of integration)
to obtain
\bea\nonumber
\Psi_1(\eps,\,t)\,-\,\psi_1(\eps,\,t)
&=&\lim_{r\to -\infty}\ -\ 
\int_r^t\,U_{\eps}(t,\,s)\
\left(\,\frac i{\eps}\,\,H(s)\,\psi_1(\eps,\,s)\,+\,
\frac{\partial\phantom{s}}{\partial s}\,\psi_1(\eps,\,s)
\,\right)
\,ds\\[5pt]
\nonumber
&=&
\frac i{\eps}\
\int_{-\infty}^t\,U_{\eps}(t,\,s)\ \zeta(\eps,\,s)
\,ds.
\eea
To bound the difference between $\Psi_1(\eps,\,t)$ and $\psi_1(\eps,\,t)$,
we use the unitarity of the evolution operator together with 
the expression 
we derived for  $\zeta(\eps,\,s)$ in Section \ref{truncation}.

This yields
\bea\nonumber
&&\Psi_1(\eps,\,t)\,-\,\psi_1(\eps,\,t)\\[5pt]
\nonumber
&=&-\ \eps^n\,\int_{-\infty}^t\,
e^{ir/(2\eps)}\ e^{\int_{-\infty}^r\,(f(s)\,
\sum_{j=1}^{n}\,g_j(s)\,\eps^j)\,ds}\
\left(\,G_n'(r)\,+\,F_n(r)\,\right)\
U_{\eps}(t,\,r)\,\Phi_2(r)\,dr\\[5pt]
\label{Error1}
&=&
-\ \eps^n\,\int_{-\infty}^t\,
e^{ir/(2\eps)}\ e^{\int_{-\infty}^r\,(f(s)\,
\sum_{j=1}^{n}\,g_j(s)\,\eps^j)\,ds}\
G_n'(r)\
U_{\eps}(t,\,r)\,\Phi_2(r)\,dr\\[5pt]
\label{Error2}
&&-\ \eps^n\,\int_{-\infty}^t\,
e^{ir/(2\eps)}\ e^{\int_{-\infty}^r\,(f(s)\,
\sum_{j=1}^{n}\,g_j(s)\,\eps^j)\,ds}\
F_n(r)\
U_{\eps}(t,\,r)\,\Phi_2(r)\,dr,
\eea
where $F_n(r)$ is $-i/\eps^{n+1}$ times the expression
(\ref{zeta1}) with the first term removed.

We now examine (\ref{Error1}), which, as we shall see,
dominates (\ref{Error2}).
Using (\ref{absintegral}), and (\ref{intsum}) we first see that
\bea\nonumber
e^{\int_0^r\,(f(s)\,\sum_{j=1}^{n}\,g_j(s)\,\eps^j)\,ds}
\ =\ 1\ +\ O(\eps),
\eea
where the $O(\eps)$ is uniform in $r$.

We also have
\bea\nonumber
U_{\eps}(t,\,r)\ \Phi_2(r)\ =\ 
e^{-i(t-r)/(2\eps)}\ \Phi_2(t)\ +\ O(\eps)\ =\ 
e^{ir/(2\eps)}\ \psi_2(\eps,\,t)\ +\ O(\eps),
\eea
where the $O(\eps)$ is uniform in $t$ and $r$.
Using these estimates together with Proposition \ref{gbounds} and 
(\ref{absintegral}), we see that (\ref{Error1}) equals
\bea\label{grandpa}
-\ \eps^n\ \psi_2(\eps,\,t)\ 
\int_{-\infty}^t\ e^{i\,r/\eps}\ G_n'(r)\ dr 
\ +\ O(\eps^{n+1}\,n!),
\eea
with the remainder estimate uniform in $t$.
By (\ref{Gexpression}), (\ref{gamma}), (\ref{beta}),
(\ref{stupid}), our choice of $n\simeq 1/\eps$, and Lemma \ref{favorite}
with $m=n+1$,
we see that for $1/2<\gamma <1$, (\ref{grandpa}) equals 
\bea\label{blah}\nonumber
&&-\ \beta^*\ \eps^n\ 
\frac{(n+1)!}{n+1}\
\sqrt{\frac{\pi}{2\,(n+1)}}\ 
\left\{\,\mbox{\rm erf}\left(\,
\sqrt{\frac {n+1}2}\ t\,\right)\,+\,1\ +\ 
O(n^{-(\gamma-1/2)})\,\right\}\psi_2(\eps,t)
\\
&&\ +\ O(\eps^{n+1}\,n!).
\eea
We deal with the prefactor before the curly bracket
using Stirling's Formula 
\bea\nonumber
k!\ =\ \sqrt{2\,\pi\,k}\ k^k\ e^{-k}\ (1+O(1/k))
\eea
(applied with $k=n+1$) 
and with the easily checked asymptotics 
\bea\nonumber
e^{x-\grintl x \grintr}
\left(\frac{\grintl x \grintr}{x}\right)^{\grintl x\grintr -1}
\ =\ 1\,+\,O(1/x)\qquad\mbox{as\ }x\ra \infty.
\eea

By the mean value theorem, the erf function satisfies
\bea
\mbox{erf }(\sqrt{x}\tau )&=&
\mbox{erf }(\sqrt{\grintl x \grintr}\,\tau )
\,+\,\frac{\tau \sqrt{y}}{2 y}\,
e^{-\tau^2 y}\,(x-\grintl x \grintr),\quad \mbox{ where } \quad
y\in ( \,\grintl x \grintr,\,x \,), \ \tau\in\R,\nonumber\\
&=&\mbox{erf }(\sqrt{\grintl x \grintr}\tau )
\,+\,O(1/x), \quad 
\mbox{ uniformly in  }\ \,\tau . 
\eea
Finally, the remainder term in (\ref{grandpa}) is of order 
$O(\sqrt{\eps}\,e^{-1/\eps})$ so that 
the quantity (\ref{Error1}) equals
\be\label{yippie}
-\ \beta^*\ \pi\ e^{-1/\eps}\ 
\left\{\,\mbox{\rm erf}\left(\,
\sqrt{\frac 1{2\,\eps}}\ t\,\right)\,+\,1
\ +\ O(\eps^{\mu})\,\right\}\psi_2(\eps, t)
\ + \ O(e^{-1/\eps}\,\sqrt{\eps}),
\ee
uniformly in $t$ for any $0<\mu<1/2$. 

Since $F_n\in L^1$, a similar analysis of
(\ref{Error2}) shows that it is 
uniformly bounded in $t$ and that its
absolute value is smaller than the estimate
of the absolute value of the
integral in (\ref{grandpa})
by a factor of $\ds C\ \frac{\sqrt{n+1}\ \log(n-2)}{n}$.
This factor arises as the product of 
$\ds\frac{\log(n-2)}{n}$ from Lemma \ref{yeah} and a factor
of $\sqrt{n+1}$ that appears in the denominator of
(\ref{answer1}). Therefore, by estimates similar to those above, 
(\ref{Error2}) is of order 
$O(e^{-1/\eps}\,\sqrt{\eps}\,\ln(1/\eps))$, as $\eps\ra 0$.
Hence, for any $0<\mu<1/2$,
\bea\label{eeha}
\Psi_1(\eps,\,t)&=&\psi_1(\eps, t)\ -\
\beta^*\ \pi\ e^{-1/\eps}\ 
\left\{\,\mbox{\rm erf}\left(\,
\sqrt{\frac 1{2\,\eps}}\ t\,\right)\,+\,1
\ +\ O(\eps^{\mu})\,\right\}\ \psi_2(\eps, t)\nonumber\\
&+&O(e^{-1/\eps}\,\sqrt{\eps}\,\ln(1/\eps)),
\eea
uniformly for $t\in\R$. 

A similar analysis of the solution $\Psi_2(\eps,\,t)$ of (\ref{SE}) that
coincides with  $\psi_2(\eps,\,t)$ as $t\ra -\infty$ leads to the estimate
\bea\label{eeha2}
\Psi_2(\eps,\,t)&=&\psi_2(\eps, t)\ +\ \beta^*\ \pi\ e^{-1/\eps}\ 
\left\{\,\mbox{\rm erf}\left(\,
\sqrt{\frac 1{2\,\eps}}\ t\,\right)\,+\,1
\ +\ O(\eps^{\mu})\,\right\}\ \psi_1(\eps, t)\nonumber\\
&+&O(e^{-1/\eps}\,\sqrt{\eps}\,\ln(1/\eps)),
\eea
uniformly for $t\in\R$. 

\vskip .5cm
For any $0<\mu<1/2$, the following estimates follow directly from
(\ref{eeha}), (\ref{eeha2}) and the unitarity 
of $U_\eps(t,r)$:
\bea\nonumber
\Psi_j(\eps,\,t)&=&\psi_j(\eps,\,t)\ +\ O(e^{-1/\eps}),\\[7pt]\nonumber
\|\psi_j(\eps,\,t)\|&=&1\ +\ O(e^{-1/\eps}\,\eps^{\mu}),
\qquad\mbox{and}\\[7pt]\nonumber
\bra\,\psi_1(\eps,\,t),\,\psi_2(\eps,\,t)\,\ket
&=&O(e^{-2/\eps}\,\eps^{\mu})
\eea
for $j=1,2$.
Thus, by (\ref{opti1}) and (\ref{opti2}), we have
\bea\nonumber
\lim_{|t|\ra \infty}|\bra \Phi_j(t),\,\psi_j(\eps, t)\ket|\ =\
\left|\,e^{(-1)^{j+1}\int_{-\infty}^{\infty} f(s) 
g(\eps, \,(-1)^{j+1}s)\,ds}\,\right|\ =\
1\ +\ O(e^{-1/\eps}\,\eps^{\mu}).
\eea

\vskip .5cm
We prove that the value of
$\beta^*$ is $1/(\pi\sqrt{2})$ by comparing 
the asymptotics above as $t\ra\infty$ with the transition 
amplitude (\ref{oldstuff}).

This proves Theorem \ref{supertheorem} with 
\bea\label{changevar}
\chi_j(\eps,t)=\psi_j(\eps, t/\delta)
\eea
when we return to the original notation 
$\eps \mapsto \eps/(E\,\delta)$ and 
original time variable $t\mapsto t/\delta$.\ep

\end{document}